# SMF-Coupled Compact Ground Terminal with Advanced Filtering Towards Daylight C-Band Satellite QKD

Argiris Ntanos[(1)], Aristeidis Stathis[(1)], Panagiotis Kourelias[(1)], Evridiki Kyriazi [(1)], Panagiotis Toumasis[(1)], Nikolaos K. Lyras[(2)], Nikolaos Makris[(3)], Sotirios Tsavdaridis [(5)], E.M. Xilouris [(6)], Athanasios Marousis[(6)], Ilias Papastamatiou [(4)], Athanasios D. Panagopoulos[(1)], Konstantinos Vyrsokinos[(5)], Kleomenis Tsiganis[(5)], George T. Kanellos[(3)], Hercules Avramopoulos[(1)] and Giannis Giannoulis[(1)]

[(1)] National Technical University of Athens, Athens, Greece ntanosargiris@mail.ntua.gr
[(2)] Optoelectronics Section, European Space Agency, Noordwijk, The Netherlands
[(3)] National Kapodistrian University of Athens, Athens, Greece
[(4)] GRNET S.A. – National Infrastructures for Research and Technology, Athens, Greece
[(5)] Aristotele University of Thessaloniki, Thessaloniki, Greece
[(6)] National Observatory of Athens, Athens, Greece

**Abstract** *We demonstrate a compact, high-isolation C-band optical ground terminal for satellite-QKD, achieving more than 120 dB daylight background suppression and 135 dB crosstalk noise isolation. Successful QKD over an outdoor 100m FSO validate its feasibility for integration of satellite QKD with urban fiber segments.* © 2025 The Author(s)

**Introduction**

Satellite-based Quantum Key Distribution (SatQKD) has been firmly established as a viable technology for enabling global-scale secure communication, following the landmark demonstrations by the Micius [1] and Jinan-1 satellites [2], which verified the feasibility of space-based QKD. However, transitioning from feasibility demonstrations to scalable, operational networks presents several critical engineering challenges: satellite platforms must be miniaturized to enable constellation deployment; ground stations must be portable and autonomous; and the entire system must support real-time, high-rate secure key exchange with minimal manual intervention. Ensuring interoperability between space-based quantum links and terrestrial fiber networks is another central design consideration for SatQKD. Operating at the 1550 nm telecom wavelength directly facilitates this integration, leveraging decades of advancements in low-loss optical components, filtering technologies, and high-performance detectors [3]. Efficient coupling of the free-space optical (FSO) beam into Single-Mode Fiber (SMF) not only ensures compatibility with existing networks but also leverages the spatial filtering properties of SMF to suppress background noise, enabling SatQKD operation even under daylight conditions [4]. From a deployment perspective, SMF coupling further allows the transfer of keys from open areas—such as rooftops, where telescopes are typically located—to trusted environments where detection equipment is shielded from environmental factors and where quantum keys can be distributed to more end users via a passive fiber network [5].

In Europe, the Eagle-1 mission aims to launch a C-band QKD terminal to enable the space segment of the EuroQCI initiative [6]. To support the Eagle-1 mission, initial technical requirements for the deployment of Optical Ground Stations (OGS) within national QCIs across Europe have been made public [7], providing early guidance for system design and implementation. Among these requirements are measures to ensure sufficient isolation between classical communication channels and quantum signals within the C-band, as well as strategies to mitigate diffuse background noise at the fiber entrance.

Motivated by emerging system requirements and the upcoming deployment of C-band SatQKD, we experimentally investigate, to the best of our knowledge for the first time, the use of spectral, spatial, and temporal filtering to simultaneously suppress background photon counts and isolate quantum signals from classical C-band communication, under both nighttime and daylight conditions, using a compact 8-inch telescope equipped with a custom Commercial-off-The-Shelf (COTS)-based fiber-coupling assembly. The developed fiber-coupling module achieved free-space-to-SMF coupling efficiencies higher than 10% under stable wireless channel conditions. The presented ground terminal integrates advanced spatial, spectral, and temporal filtering, achieving over 120 dB suppression of crosstalk and out-of-band noise. The filtering system's performance is further validated through background noise measurements under daylight conditions, achieving only 50 counts per second (cps) and 150 cps when the telescope is pointed at a solar elongation angle of 45° and 20°, respectively. Moreover, we report the successful transmission of QKD signals over the developed small-sized ground terminal, confirming the system's ability to handle qubits with negligible Quantum Bit Error Rate (QBER) penalty. These results demonstrate the potential for deployable, scalable 1550 nm SatQKD ground stations, showing that fiber-integrated, modular terminal architectures with high-isolation filtering can meet the operational demands of upcoming space-QKD missions.

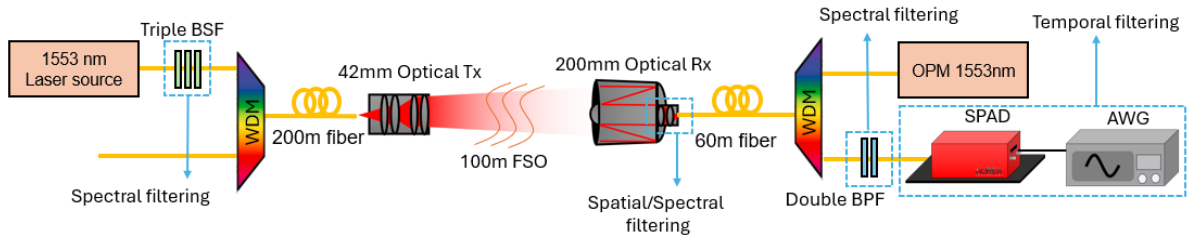

**Figure 1.** Experimental setup for assessing the performance of background and crosstalk noise reduction.

**Results and Discussion**

SatQKD requires a different approach regarding noise suppression compared to fiber-based QKD, where quantum signals can be transmitted over a dark fiber. In future satellite-based QKD systems, quantum signals are expected to be transmitted alongside classical signals at neighbouring wavelengths, used for synchronization and post-processing communication. Additionally, to enable SatQKD operations during both nighttime and daylight hours, effective suppression of background solar radiance is essential. We propose and evaluate a filtering scheme designed to supress both crosstalk and diffuse background sky radiance. Figure 1 presents the experimental setup used for noise suppression between the classical signal centered at 1553.33 nm and the quantum passband centered at 1550.12 nm. A 20 cm optical telescope has been used as the optical ground terminal, with a custom SMF coupling module developed and mounted at the back of the telescope. To evaluate the coupling efficiency, a 100m FSO link has been established, where 42.5 mm air-spaced achromatic fiber collimator served as the optical transmitter. The signals are directed to the rooftop FSO terminals and then routed back into the laboratory using two dedicated deployed optical fibers of 60 and 200 meters with an Insertion Loss (IL) of 0.5 dB and 1 dB, respectively. Under the relatively stable conditions of this configuration—where atmospheric turbulence is negligible, more than ~10% free-space-to-SMF coupling efficiency was achieved. The total attenuation introduced by the FSO channel, was measured to be approximately 13 dB, with 3 dB attributed to reflectivity losses from the telescope's optical coatings, 1 dB to the fiber-coupling optics, and the remaining losses due to free-space-to-SMF-coupling process. To mitigate crosstalk, a triple-stage 0.8 nm band-stop filter with an IL of 1.5 dB is applied to the classical signal prior to transmission, directly suppressing crosstalk noise at the quantum passband. The combination and separation of the classical and quantum signals is performed with two WDM modules (IL=1.5 dB) are utilized. On the receiver side an additional two stage 0.2nm passband filter with 3.2 dB IL. The passband filter that is employed is able to supress both the background sky radiance noise as well as any remaining crosstalk noise from the classical signal. Finally, temporal filtering has been used, by operating the Single Photon avalanche Detector (SPAD) in gating mode to further suppresses any remaining noise. For the noise measurement, the SPAD was operating at a quantum efficiency of 10%.

Figure 2 illustrates the suppression performance achieved through the combination of spectral, spatial, and temporal filtering in the proposed setup. First, regarding crosstalk noise suppression, the configuration reduced classical channel leakage by a total of 121 dB. Specifically, 101 dB suppression was achieved by the combination of the notch and the band-pass spectral filtering, and an additional 20 dB by temporal filtering, implemented through SPAD gating with 1% duty cycle signal.

Regarding the suppression of background sky radiance, it was observed that nighttime background noise was practically suppressed to zero, without the need of optical filters due to the spatial filtering enabled by SMF coupling. On the other hand, during daylight operation narrow passband filtering is required. After the application of all three filtering mechanisms the background solar radiance noise has been supressed to a value of 136 dB. As a result, with the telescope at a solar elongation angle of ~45 degrees the overall noise coupling from the 20cm and after applying the previous spectral and temporal filtering, the noise click rate in daylight was suppressed down to just 50 cps. Finally, the value of the background noise was found to be dependent to the solar elongation angle, where the noise increased up to 3 times when the telescope pointed at a maximum angle of ~20 degrees away from the sun, resulting in 150 noise cps in gated mode, whereas for elongation angles greater than 70 degrees the background noise dropped practically to zero. It was also observed that during daytime a large amount of noise directly coupled into the 60m SMF segment. After covering the exposed fiber parts and applying spectral and temporal filtering this noise was able to be supressed down to around 2 cps. It should be noted here that since in our configuration the receiver's field of view is constrained by the size of the fiber mode field diameter the detection rate of diffuse background photons coupled into such a system becomes almost independent of the receiver optical system parameters [8]. Finally, Figure 3 presents the measured noise click rates under clear sky and cloudy sky conditions, recorded in free-running mode. It is evident that, in regions of the Cloud-Free-Line-of-sight (CFLOS) area, the noise level is effectively reduced to the dark count rate (DCR) baseline (~500 cps). In contrast, areas with scattered clouds exhibit an increase in noise by up to an order of magnitude.

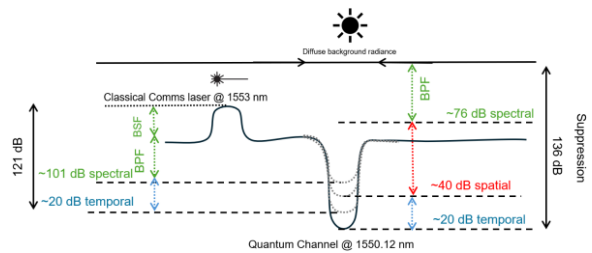

**Figure 2.** Spectral, spatial, and temporal filtering used to suppress background noise and classical communication laser leakage in the quantum channel.

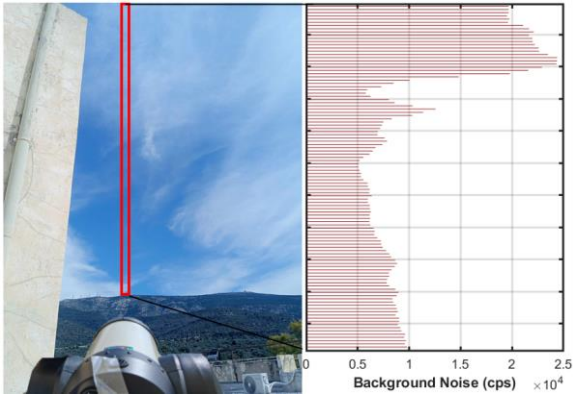

**Figure 3.** Background Sky radiance noise counts

To validate the ability of the developed optical terminal to support QKD communication the transmission of phase-encoded quantum signals over the developed 100m FSO link has been demonstrated. Figure 4a presents the experimental configuration for the FSO-QKD link employing the Clavis XG QKD system [4]. The quantum signal operates at a wavelength of 1551.72 nm (ITU-T Channel 32). In this configuration the classical communication between Alice and Bob is facilitated via a dedicated back-to-back fiber connection. As shown in Figure 4b, the SKR, visibility, and QBER were monitored continuously over a period of 1200 seconds. The SKR remained highly stable around 4.2 kbps, visibility stayed close to 98%, and the QBER remained below 1%, confirming stable key generation, high quantum interference quality, and robust optical alignment throughout the measurement period. It should be noted, however, that the experimental conditions were relatively stable, with negligible atmospheric turbulence, whereas actual satellite links introduce additional impairments such as beam wander, scintillation, Doppler shifts, and pointing inaccuracies. These factors would likely impose further challenges, requiring adaptive optics and more advanced tracking systems to maintain similar performance in orbit-ground QKD scenarios and some of these were extensively discussed in [9].

**Conclusions**

We presented the first experimental validation of quantum signal integration alongside neighbouring service wavelengths for C-band SatQKD. Spatial, spectral, and temporal filtering achieved >120 dB crosstalk suppression and > 135 dB daylight noise reduction. The developed compact 8-inch fiber-coupled optical ground terminal achieved >10% coupling efficient along with stable key generation over a 100 m outdoor FSO link validating the feasibility of small-sized optical ground terminals for future satellite QKD networks and urban quantum communication infrastructures.

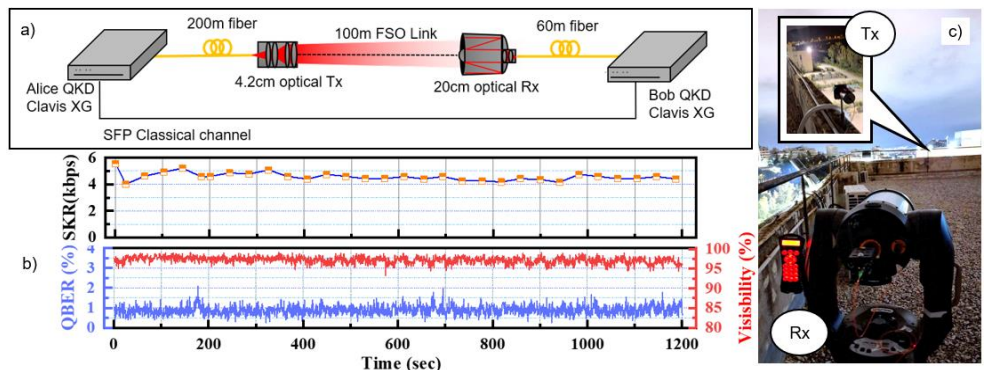

**Figure 4.** a) FSO-QKD experimental setup, b) SKR (kbps), Visibility (%) and QBER (%) values over time c) Transmitting and receiving optical terminals of the 100 m FSO link

**Acknowledgements**
This work has received funding from the European Union's Horizon Europe Research and Innovation programme under the project "Quantum Security Networks Partnership" (QSNP, Grant Agreement No 101114043). This work has been supported by the project HellasQCI, under grant agreementNo.101091504. Views and opinions expressed are, however, those of the authors only and do not necessarily reflect those of the European Union. The European Union cannot be held responsible for them. This work has also been supported by European Space Agency under Contract No. 4000145410/24/UK/AL, entitled "Holomondas Observatory Upgrade for Optical and Quantum Communication".